\documentclass{article}

\usepackage{arxiv}

\usepackage[utf8]{inputenc} % allow utf-8 input
\usepackage[T1]{fontenc}    % use 8-bit T1 fonts
\usepackage{hyperref}       % hyperlinks
\usepackage{url}            % simple URL typesetting
\usepackage{booktabs}       % professional-quality tables
\usepackage{amsfonts}       % blackboard math symbols
\usepackage{nicefrac}       % compact symbols for 1/2, etc.
\usepackage{microtype}      % microtypography
\usepackage{lipsum}		% Can be removed after putting your text content
\usepackage{graphicx}
\usepackage{float}

\graphicspath{{Figures/}} %Setting the graphicspath

\title{Fully automated analysis of muscle architecture from B-mode ultrasound images with deep learning}

\author{ \href{https://orcid.org/0000-0002-5332-1188}{\includegraphics[scale=0.06]{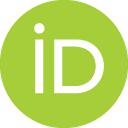}\hspace{1mm}Neil J.~Cronin}\\
	$^1$Faculty of Sport and Health Sciences\\
	University of Jyväskylä, Finland,\\
	$^2$School of Sport and Exercise\\
	University of Gloucestershire, UK\\
	\texttt{neil.j.cronin@jyu.fi} \\
	\And
	\href{https://orcid.org/0000-0002-7697-2813}{\includegraphics[scale=0.06]{orcid.png}\hspace{1mm}Taija~Finni} \\
	Faculty of Sport and Health Sciences\\
	University of Jyväskylä, Finland\\
	\texttt{taija.juutinen@jyu.fi} \\
	\And
	\href{https://orcid.org/0000-0002-1289-246X}{\includegraphics[scale=0.06]{orcid.png}\hspace{1mm}Olivier~Seynnes} \\
	Department for Physical Performance\\
	Norwegian School of Sport Sciences, Oslo, Norway\\
	\texttt{olivier.seynnes@nih.no} \\
}

\renewcommand{\shorttitle}{Automated ultrasound analysis with deep learning}

\hypersetup{
pdftitle={Fully automated analysis of muscle architecture from B-mode ultrasound images with deep learning},
pdfsubject={computer vision, deep learning},
pdfauthor={Neil J.~Cronin, Taija~Finni, Olivier~Seynnes},
pdfkeywords={ultrasound, U-net, convolutional neural network, muscle architecture},
}

\begin{document}
\maketitle

\begin{abstract}
	B-mode ultrasound is commonly used to image musculoskeletal tissues, but one major bottleneck is data interpretation, and analyses of muscle thickness, pennation angle and fascicle length are often still performed manually. In this study we trained deep neural networks (based on U-net) to detect muscle fascicles and aponeuroses using a set of labelled musculoskeletal ultrasound images. We then compared neural network predictions on new, unseen images to those obtained via manual analysis and two existing semi/automated analysis approaches (SMA and Ultratrack). With a GPU, inference time for a single image with the new approach was around 0.7s, compared to 4.6s with a CPU. Our method detects the locations of the superficial and deep aponeuroses, as well as multiple fascicle fragments per image. For single images, the method gave similar results to those produced by a non-trainable automated method (SMA; mean difference in fascicle length: 1.1 mm) or human manual analysis (mean difference: 2.1 mm). Between-method differences in pennation angle were within 1º, and mean differences in muscle thickness were less than 0.2 mm. Similarly, for videos, there was strong overlap between the results produced with Ultratrack and our method, with a mean ICC of 0.73, despite the fact that the analysed trials included hundreds of frames. Our method is fully automated and open source, and can estimate fascicle length, pennation angle and muscle thickness from single images or videos, as well as from multiple superficial muscles. We also provide all necessary code and training data for custom model development. 
\end{abstract}

% keywords can be removed

\section{Introduction}

B-mode ultrasound is commonly used to image musculoskeletal tissues, and numerous studies have used this method to examine muscle function both statically and dynamically \cite{Cronin2013,Seynnes2015}. Since it is generally much cheaper and more portable than other imaging modalities (e.g. MRI), this method can be used across a wide range of domains. However, one major bottleneck associated with ultrasound is the interpretation of data. To date, analyses of muscle thickness, pennation angle and fascicle length are most often performed manually. 

In recent years, attempts have been made to automate or semi-automate at least parts of the analysis process, making it faster and potentially more objective \cite{Cronin2011c,Rana2009,Marzilger2018,Drazan2019,Farris2016}. A number of papers have also reported full automation of some muscle architectural parameters \cite{Caresio2017,Zhou2018,Seynnes2020,Darby2013}. However, existing automated methods are associated with several drawbacks. Firstly, the majority of ‘automated’ methods are actually only semi-automated, requiring some user input to delineate some of the features or regions of interest, or to adjust certain pre-processing steps. Secondly, most methods rely on an effective pre-processing of images to enhance the contrast of structures of interest. This step may fail when automated, if image parameters do not suit the designed filters, or when it is semi-automated, if the filters are adjusted improperly. Thirdly, most software implementations in this area are not open source, effectively limiting their assessment and their uptake in the broader community. Clearly, there is a need for a more versatile tool that is accessible to all users, and enables a range of parameters to be analysed.

The past decade has brought huge advances in the field of artificial intelligence. Specifically, in relation to medical imaging, convolutional neural networks have shown great promise to identify features in images from a range of modalities \cite{Tajbakhsh2016,Milletari2016}, including ultrasound \cite{Cunningham2018,Cunningham2020}. Such an ‘intelligent’ approach can be trained on a set of labelled images, and can effectively ‘learn’ to identify relevant features in new, previously unseen images. Cunningham et al. \cite{Cunningham2018} were the first to use convolutional neural networks to analyse muscle architecture. Although their method was able to detect fascicle trajectories, they reported errors in pennation angle as high as 6º. This may have been due to a relatively small training dataset, and/or an insufficiently robust neural network approach. Recently, a method designed specifically for medical image analysis (U-net; \cite{Ronneberger2015}) was introduced, and was found to outperform existing approaches at various medical image segmentation tasks (albeit not related to ultrasound imaging). This method makes extensive use of image augmentation during training, which helps to compensate for relatively small labelled datasets and can improve model robustness. 

In this study we built an implementation based on U-net, and trained it to detect muscle fascicles and aponeuroses using a set of labelled musculoskeletal ultrasound images. We demonstrate that the approach can reliably identify these structures in previously unseen scans of different lower limb muscles. Moreover, we are able to compute several relevant metrics for muscle architecture studies (fascicle length, pennation angle, muscle thickness) including the detection of multiple fascicles, and to do so for single images and videos. Most importantly, our approach yields results that are comparable to those obtained with existing analysis methods, whilst also overcoming some of the limitations of such methods. We also provide the full Python code and training datasets to allow others to use our trained algorithm, and/or to build upon our approach. 

\section{Methods}

\paragraph{Experimental approach.}We trained two neural networks to identify features from musculoskeletal images. One model was trained to identify aponeuroses, while the other was trained to detect muscle fascicles. We used a supervised learning approach, which means that for each ultrasound image, we also provided the neural networks with a label, which highlighted the regions that we wanted to identify. The image-label pairs served as the inputs to the model.

\paragraph{Data.}We compiled a large volume of anonymised single image and video data obtained from different muscles (medial and lateral gastrocnemius, vastus lateralis, tibialis anterior) and with 4 different ultrasound devices, as well as from different human populations (athletes, older people, young healthy individuals) and different movements and contraction types. Individual frames were extracted from this dataset at random using a custom-written function in Python (Python Software Foundation, v3.6), resulting in a set of around 570 images for the aponeurosis model, and 310 images for the muscle fascicle models. All ultrasound data were acquired in previous studies by the authors, all of which received ethical approval from the relevant committees. The images have all been anonymised and can be found from the project repository (\url{https://github.com/njcronin/DL_Track}). 

\begin{figure}[h!]
  \includegraphics[width=\linewidth]{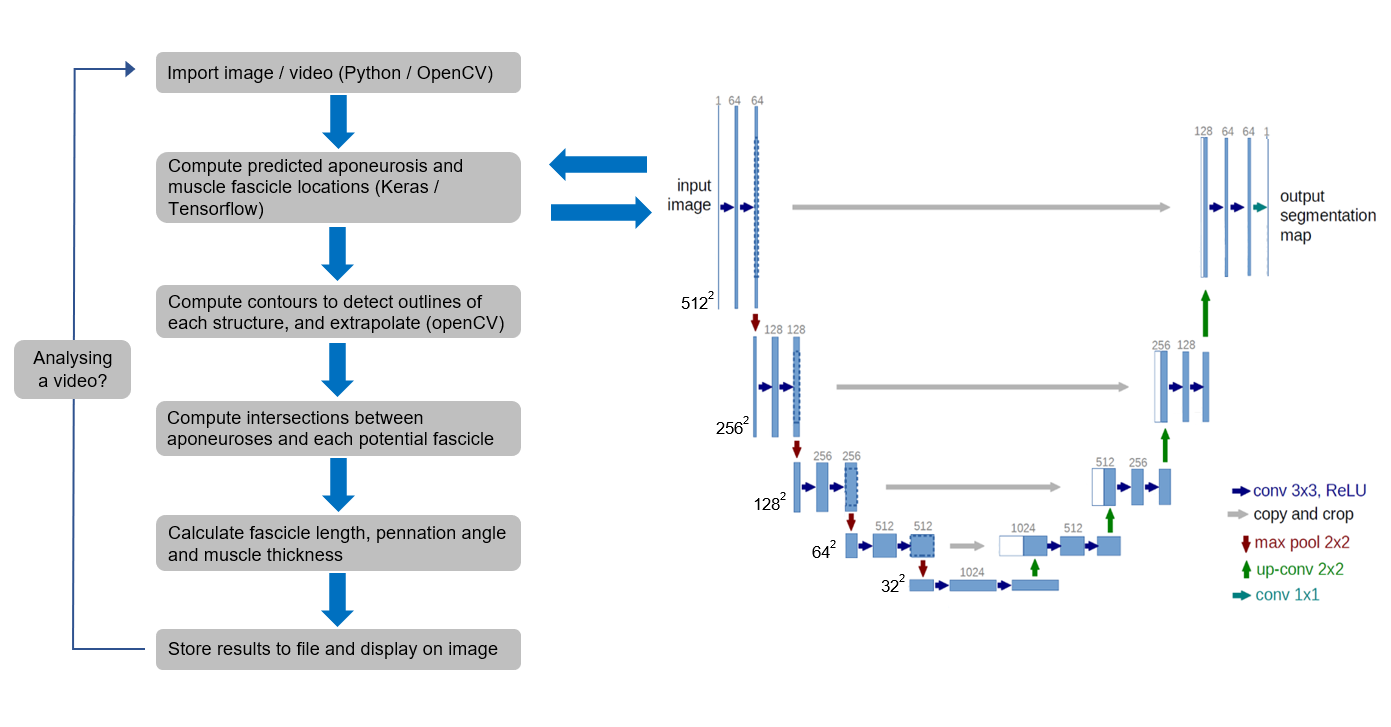}
  \caption{Schematic of the workflow, including details of the U-net model architecture (modified from \cite{Ronneberger2015}). The process is essentially identical when analysing single images or videos, since each frame is treated independently. Blue boxes represent multi-channel feature maps and the number of channels is denoted on top of each box. The x-y sizes are shown at the lower left edge of each box on the contracting side, and these are identical for the expanding side. White boxes represent copied feature maps from the contracting side, which are concatenated with those from the expanding side. Arrows denote the different operations.}
  \label{fig:Fig1}
\end{figure}

\paragraph{Neural network architecture: U-net.}The U-net neural network architecture \cite{Ronneberger2015} consists of a contracting path and an expansive path. The contracting path follows the typical architecture of a convolutional network. It consists of the repeated application of two 3x3 (unpadded) convolutions, each followed by a rectified linear unit (ReLU) and a 2x2 max pooling operation with stride 2 for downsampling. At each downsampling step, the number of feature channels is doubled. Every step in the expansive path consists of an upsampling of the feature map followed by a 2x2 up-convolution that halves the number of feature channels, a concatenation with the correspondingly cropped feature map from the contracting path, and two 3x3 convolutions, each followed by a ReLU. The cropping is necessary due to the loss of border pixels with each convolution. At the final layer a 1x1 convolution is used to map each 64-component feature vector to the desired number of classes (2 in this case). In total the network has 23 convolutional layers (see Figure ~\ref{fig:Fig1}). The output of the model is a pixelwise binary label, i.e. every pixel of an image is predicted to belong to one of two possible classes. In our approach these classes were aponeurosis/not aponeurosis (aponeurosis model), and fascicle/not a fascicle (fascicle model).

By the standards of typical deep learning applications, medical imaging is generally limited by relatively small labelled training datasets. With the current approach, this is countered by using excessive data augmentation in the form of elastic deformations of the training images. This allows the network to learn invariance to such deformations, without the need to manually label the transformed images. This is particularly important in biomedical segmentation because the tissues do undergo deformation during muscle contraction, and realistic deformations can be simulated efficiently.

\begin{figure}[h!]
  \includegraphics[width=\linewidth]{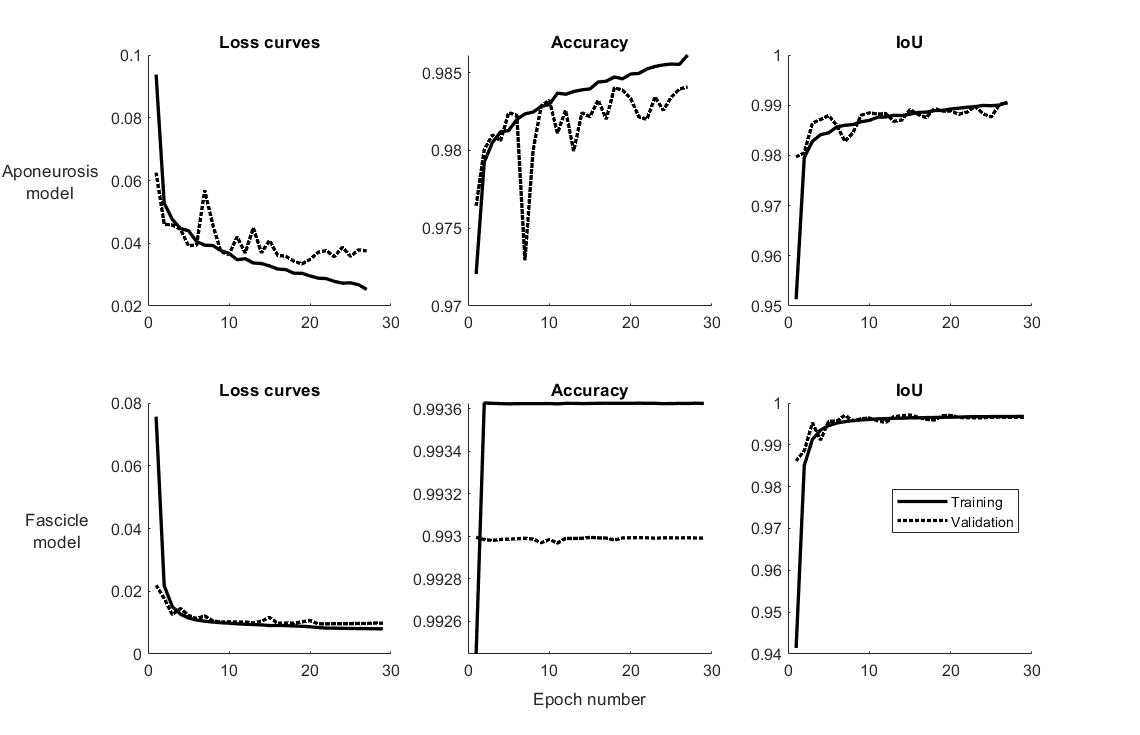}
  \caption{Loss function and intersection-over-union (IoU) training curves for the two trained models.}
  \label{fig:FigS1}
\end{figure}

\paragraph{Neural network training.}We trained two separate models, each using the same U-net architecture. Images were imported and resized to 512*512 pixels for training. In general, neural networks perform faster with smaller images, but in this case we chose the largest possible image size given RAM limitations, since the quality of ultrasound images is typically quite low, and further reductions in spatial resolution due to image downsampling would likely compromise the ability to successfully train a neural network for pixelwise labelling. We used a 90/10\% training/validation data split. Training was performed using an RTX2070 GPU and took less than one hour per model using a maximum of 50 epochs and a batch size of 1, with Adam optimizer and the binary cross-entropy loss function. Training was stopped early when overfitting was evident, as characterised by a decrease in the training error and a concomitant stagnation or increase in the test error (see Figure ~\ref{fig:FigS1}). The code runs in Python and uses a Keras frontend with a Tensorflow backend. The code and training data from this project are freely available from Github, and installation instructions are provided: \url{https://github.com/njcronin/DL_Track}. For those who do not have access to their own GPU, we also provide a Google colaboratory version, which includes free GPU access.

\begin{figure}[h!]
  \includegraphics[width=\linewidth]{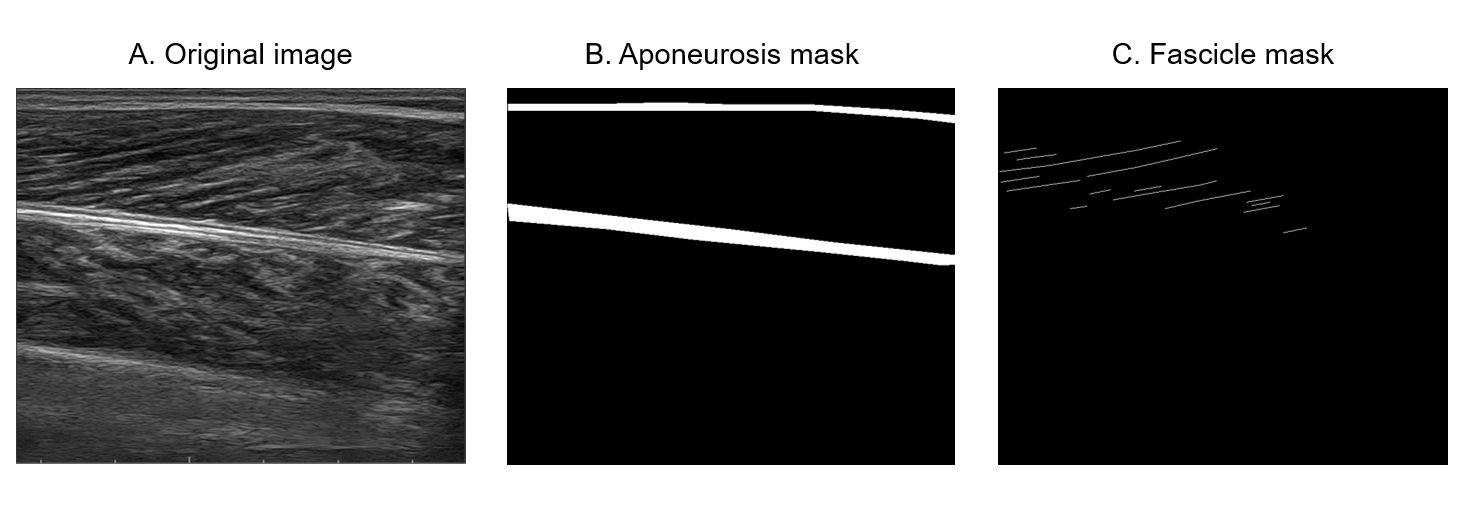}
  \caption{An example of an original ultrasound image (A) and corresponding binary masks that were used to train neural networks to identify aponeuroses (B) and muscle fascicles (C).}
  \label{fig:Fig2}
\end{figure}

For each of the aponeurosis training images, one of the researchers manually identified all instances of aponeuroses using the polygon tool in Fiji software \cite{Schindelin2012} to create a binary mask, whereby individual pixels belonging to an aponeurosis were white, and all other pixels were black. This process was repeated separately for the fascicle training set, where all instances of muscle fascicles (or parts of fascicles) were identified (1-20 per image). Along with the corresponding images, the binary masks were used as ground truth labels to train the two deep neural networks (see Figure ~\ref{fig:Fig2}).

\paragraph{Post-processing.}After processing a new image or video with the trained neural networks to identify aponeuroses and muscle fascicles, the following steps are taken to obtain relevant outcome measurements. Aponeuroses below a user-defined threshold length are removed, and where necessary, those that satisfy the length constraint are extrapolated laterally, since this can assist in finding the intersection with muscle fascicles. The trained fascicle model identifies visible fragments of fascicles. Those parts that are beyond a threshold length are extrapolated proximally and distally using a 1st order polynomial fitted to the identified structure. The intersection points between aponeuroses and fascicles are identified, and fascicle length is determined. Pennation angle is computed between each fascicle and the local slope of the lower aponeurosis (50-pixel region starting from the point of fascicle intersection). As multiple fascicle fragments are usually detected per image, data for all fragments are retained, including the image x-axis coordinates of the start and end points. Muscle thickness is determined from the central portion of the image, as the shortest distance between the superficial and deep aponeuroses.

\paragraph{Analysis metrics.}To determine the overlap between manually created aponeurosis/fascicle labels from the training set and the labels predicted by the neural network, we used a custom implementation of intersection over union (IoU). A set of 35 single test images unseen during training represented a test set, all of which were processed using the trained networks, to estimate muscle thickness, muscle fascicle length and pennation angle. The same test set and parameters were also analysed manually by all 3 authors using Fiji software, as well as using a non-trainable method of segmentation designed to analyse single images (SMA, \cite{Seynnes2020}). Comparisons between the human- and computer-generated results for this test set were done using Bland-Altman plots. For a set of videos comprising a range of tasks (e.g. passive joint rotation, maximal voluntary contraction, walking), we also compared the results of our method (referred to as DL for short) with those of Ultratrack \cite{Cronin2011c,Farris2016,Gillett2013}, which is arguably the most commonly used semi-automated approach for analysing ultrasound videos. When tracking with Ultratrack, we used keyframe correction as appropriate, and made manual corrections where necessary, so that the resulting values would act as a ‘gold standard’ for this comparison. To provide a broad comparison between our approach and Ultratrack, we computed ICCs (2,1; \cite{Salarian2016}) for a set of videos in Matlab (v2019b, The MathWorks, Inc., Natick, Massachusetts, United States). All statistical comparisons were also done using Matlab.

\section{Results}

\begin{figure}[h!]
  \includegraphics[width=\linewidth]{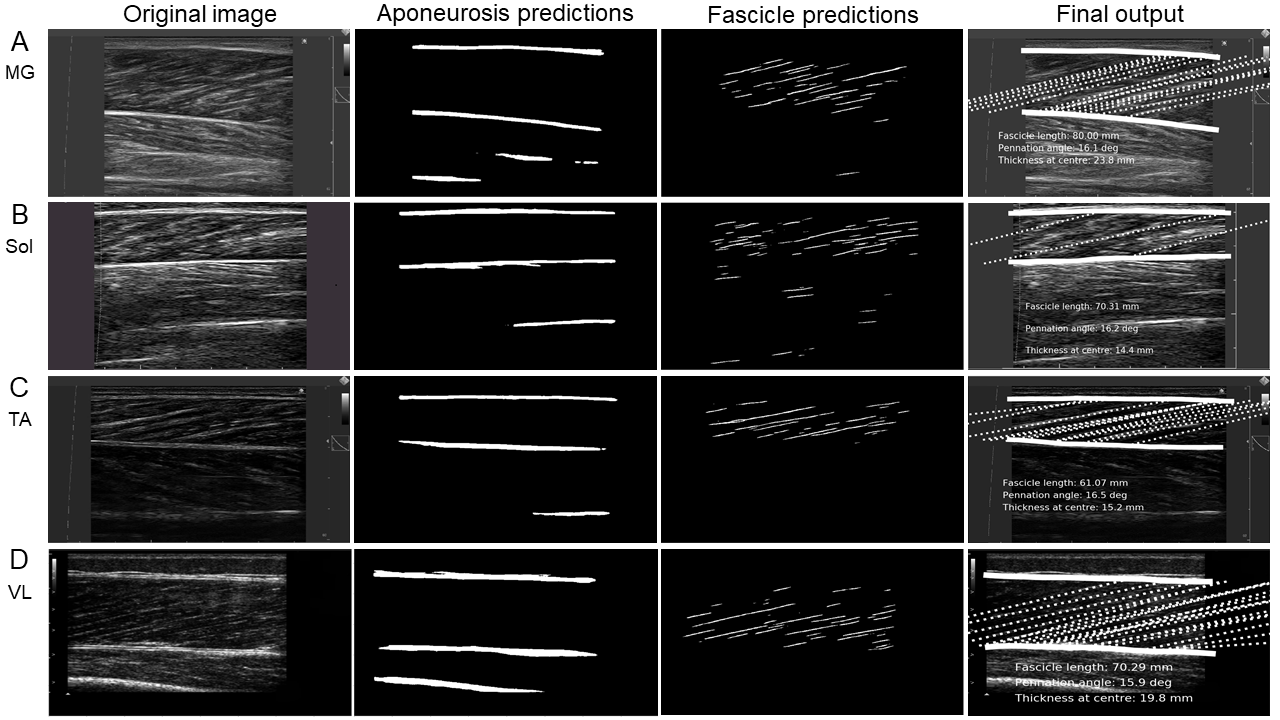}
  \caption{Examples of images annotated by the trained neural networks. These images are from the medial gastrocnemius (A), soleus (B), tibialis anterior (C) and vastus lateralis (D) muscles. Original images are shown in the leftmost column. The second and third columns show the initial neural network predictions for aponeuroses and fascicles respectively. The right column shows the final output after post-processing. Fascicle length and pennation angle results shown in the right column are the mean of all detected fascicles.}
  \label{fig:Fig3}
\end{figure}

Training of neural networks required 18-21 epochs (see Figure ~\ref{fig:FigS1}). When analysing new data, analysis time for a single image with a CPU was around 4.6s, compared to 0.7s with GPU, although these values will vary depending on specific hardware. Examples of the output provided by the trained neural networks, as well as the results of post-processing, are shown in Figure ~\ref{fig:Fig3}. These examples demonstrate that a variable number of fascicles are detected in each image.

\begin{figure}[H]
  \includegraphics[width=\linewidth]{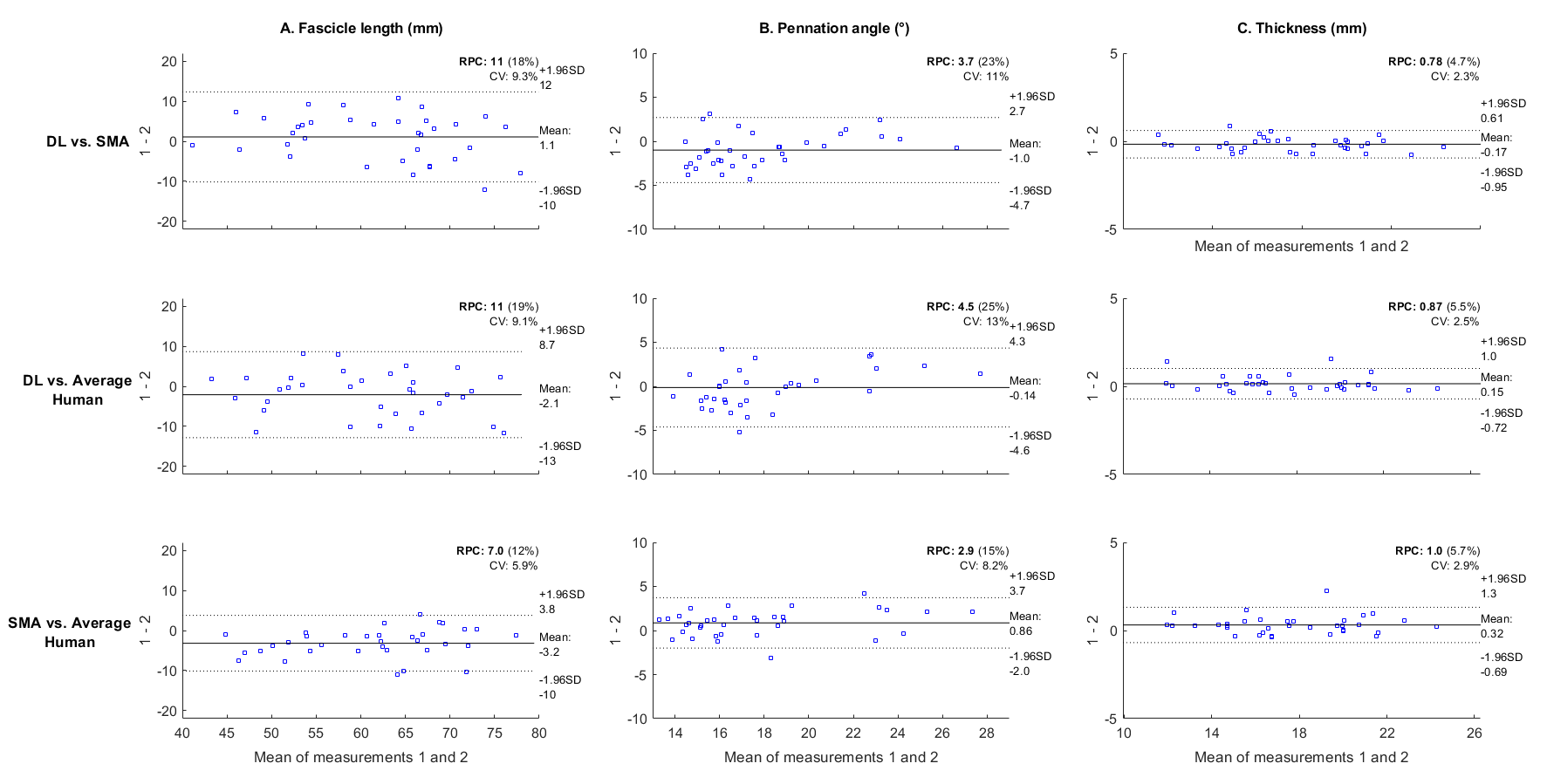}
  \caption{Bland-Altman plots of the results obtained with our approach versus the results of manual analyses by the authors (mean of all 3), as well as the SMA automated method. Results are shown for each parameter of interest: muscle fascicle length (A), pennation angle (B), and muscle thickness (C). For these plots, only the median fascicle values from the deep learning approach were used, and thickness was computed from the centre of the image. Solid and dotted lines depict bias and 95\% limits of agreement, respectively.}
  \label{fig:Fig4}
\end{figure}

\begin{figure}[h!]
  \includegraphics[width=\linewidth]{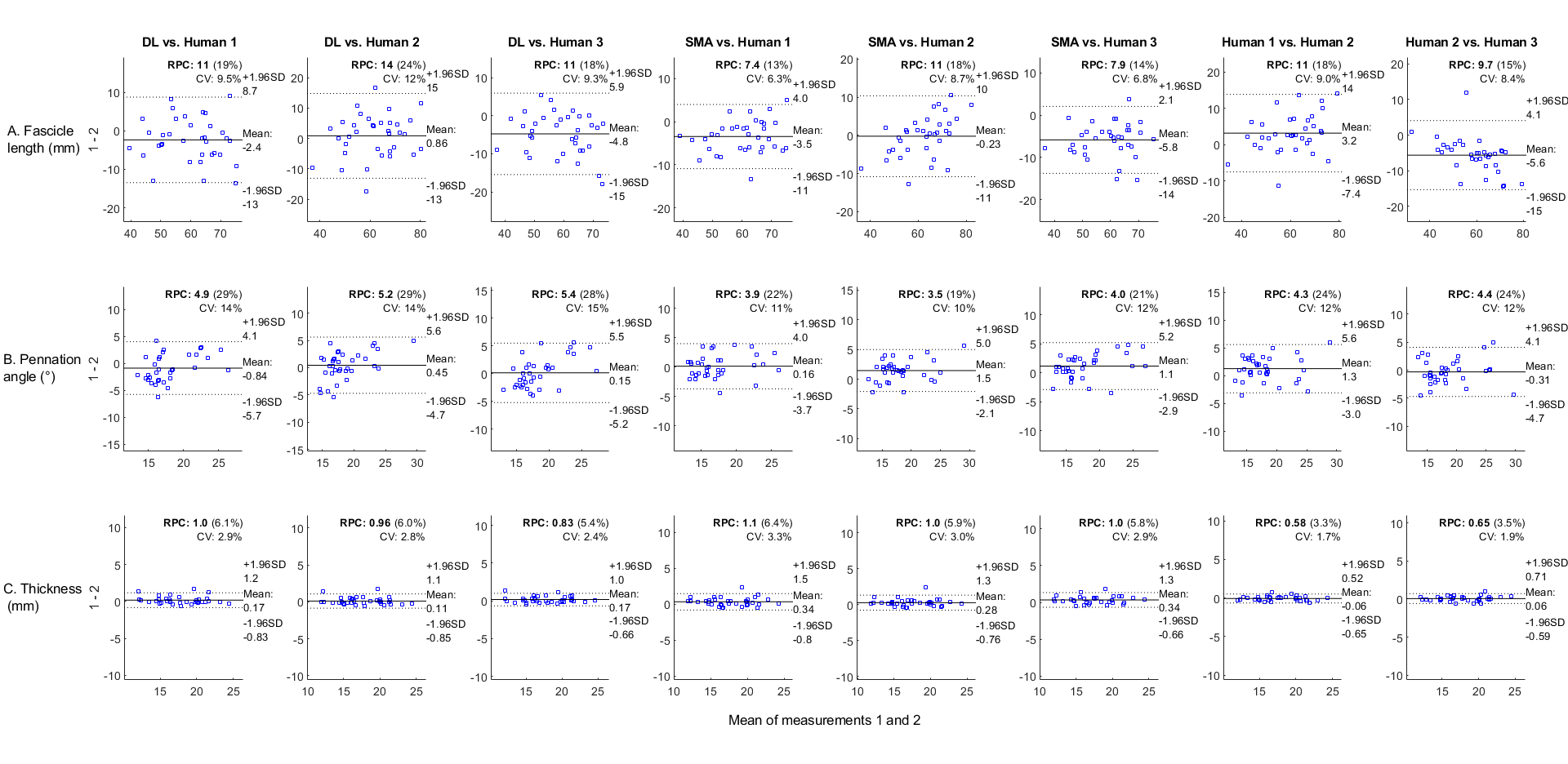}
  \caption{Bland-Altman plots of the results obtained with our approach versus the results of manual analyses by the authors (each author’s data shown separately), as well as the SMA automated method. Results are shown for each parameter of interest: muscle fascicle length (A), pennation angle (B), and muscle thickness (C). For these plots, only the median fascicle values from the deep learning approach were used, and thickness was computed from the centre of the image. Solid and dotted lines depict bias and 95\% limits of agreement, respectively.}
  \label{fig:FigS2}
\end{figure}

\begin{figure}[h!]
  \includegraphics[keepaspectratio=true, scale=0.7]{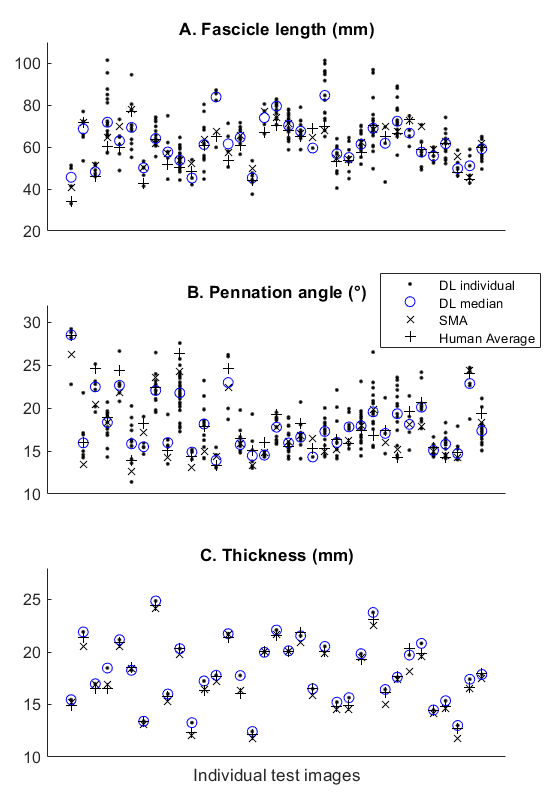}
  \caption{Analysis of the same test set of 35 images used in Figure ~\ref{fig:Fig4}, but here each individual fascicle detected by the DL method is shown (instead of just the median, as used in Figure ~\ref{fig:Fig4}), alongside the average human and SMA results.}
  \label{fig:Fig5}
\end{figure}

Figure ~\ref{fig:Fig4} shows Bland-Altman plots comparing the DL method with the non-trainable automated method (SMA), as well as with the manual analyses (average measurements from the 3 authors; for individual data see Figure  ~\ref{fig:FigS2}). Across the set of 35 test images, the mean differences between DL and SMA were: 1.1 mm (fascicle length; unpaired t-test p value: 0.879), 1.0º (pennation angle; p = 0.383), and 0.2 mm (thickness; p = 0.351). The corresponding values comparing DL with the average human results were: 2.1 mm (p = 0.249), 0.1º (p = 0.842) and 0.1 mm (p = 0.606) respectively. For the SMA method versus the average human, the values were: 3.2 mm (p = 0.170), 0.9º (p = 0.340) and 0.3 mm (p = 0.677) respectively. Figure ~\ref{fig:Fig5} shows results for the same test set of 35 images, but in this case data from all fascicles detected by DL are shown, rather than just the median. Of the 35 images, the mean values of SMA and the average human results fell within the range of individual DL values in 32 cases for fascicle length and 31 cases for pennation angle (not computed for thickness because only one value is returned per image).

\begin{figure}[!htp]
  \includegraphics[width=\linewidth]{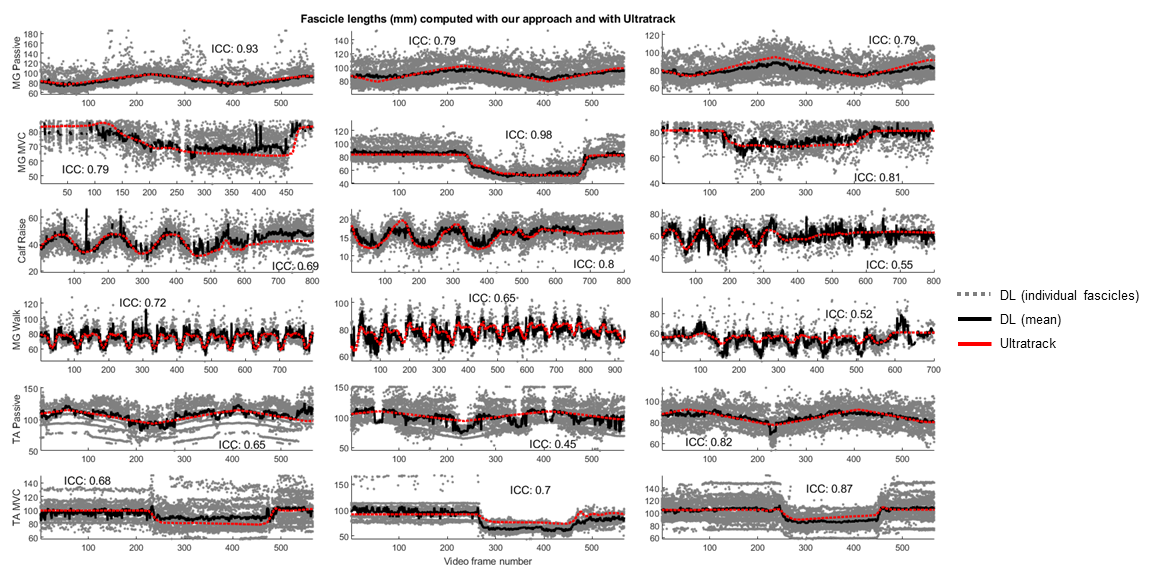}
  \caption{A comparison of fascicle lengths computed using our approach with those from UltraTrack, a semi-automated method of identifying muscle fascicles. Each row shows trials from a particular task (3 examples per task from different individuals, shown in separate columns). For our approach, the length of each individual fascicle detected in every frame is denoted by a gray dot. Solid black lines denote the mean length of all detected fascicles. Red dashed lines show the results of tracking a single fascicle with Ultratrack.}
  \label{fig:Fig6}
\end{figure}

\begin{figure}[h!]
  \includegraphics[width=\linewidth]{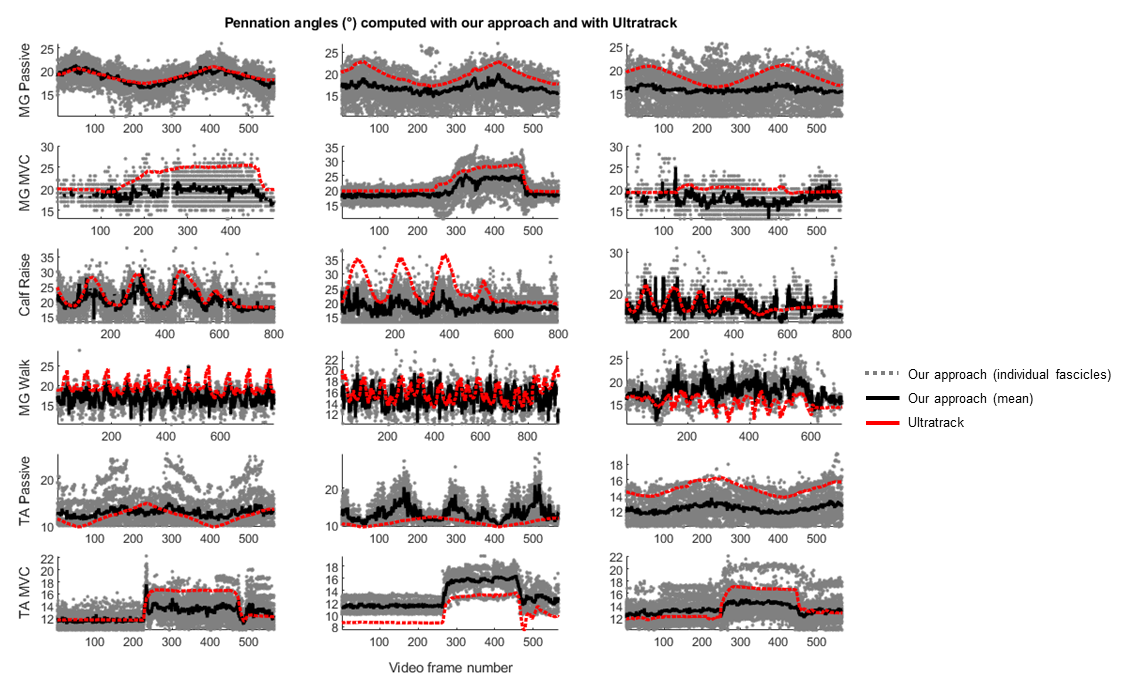}
  \caption{A comparison of our approach with UltraTrack for calculating pennation angle. The trials and tasks correspond with those shown in Figure ~\ref{fig:Fig6}.}
  \label{fig:FigS3}
\end{figure}

Figure ~\ref{fig:Fig6} shows the results of analysing fascicle length for 6 different muscle and task combinations with the DL method (3 trials per condition) and with Ultratrack (the corresponding comparison of pennation angle is shown in Figure ~\ref{fig:FigS3}). In each panel, which represents one trial, all fascicles detected by our approach are shown as individual dots, and the mean of all detected fascicles is also shown to make visual comparisons with Ultratrack easier. The number of muscle fascicles detected varied both within- and between trials/tasks. For example, the mean number of detected fascicles for the 6 tasks shown in Figure ~\ref{fig:Fig6} were: 19.5 (MG passive), 9.5 (MG MVC), 7.4 (calf raise), 3.2 (MG walk), 12.9 (TA passive), and 14.2 (TA MVC). The effects of this variation can be seen in the mean traces shown in Figure ~\ref{fig:Fig6}. The mean absolute differences between the mean traces produced by our method and the Ultratrack traces were: 3.0 mm (MG passive), 3.3 mm (MG MVC), 2.9 mm (calf raise), 5.2 mm (MG walk), 4.9 mm (TA passive) and 6.0 mm (TA MVC). ICC values calculated for each panel in Figure ~\ref{fig:Fig6} ranged between 0.45 and 0.98. Examples of tracked video output with our method can be seen in Supplementary video 1: \url{https://github.com/njcronin/DL_Track/blob/master/Supp_video_1.mp4}. 

\section{Discussion}

We present a deep learning approach for automating the analysis of muscle architecture from B-mode ultrasound images. The method is open-source, and we provide the trained models, Jupyter notebooks for analysing new images/videos, as well as the code and labelled data used to train the neural network models, allowing users to train their own. Our method detects the locations of the superficial and deep aponeuroses, as well as multiple fascicle fragments per image. We found that for single images, the method gave results that did not statistically differ from those produced by a non-trainable automated method or manual analysis by human researchers. Similarly, for videos, there was strong overlap between the results produced with Ultratrack and our method, with a mean ICC across the 18 trials in Figure ~\ref{fig:Fig6} of 0.73, despite the fact that each of these trials included hundreds of frames.

\subsection{Analysis of single images}In some study designs, it is desirable to compute architectural parameters from single images, e.g. those taken from the same individual before and after an intervention, or at different stages of the ageing process. Here we compared our approach to our own existing SMA method \cite{Seynnes2020} (which was primarily designed to process individual images), as well as to manual analysis, which is still arguably the most common method of computing architectural parameters. Despite the fact that the DL method detects multiple fascicles per frame across the width of the image, comparing the median value from this analysis with the other approaches generally revealed similar results for all parameters. For fascicle length, the mean difference between DL and other methods was 1.1 mm (vs. SMA) and 2.1 mm (vs. humans), although peak absolute differences were around 12 mm. For pennation angle, mean differences between DL and other methods were between 0 and 1º, and mean differences in muscle thickness were less than 0.2 mm for the DL comparisons.

As noted above, the DL approach is designed to allow multiple fascicles to be detected. In order to enable the above-mentioned comparisons, we used the median value of all detected fascicles, since the SMA method only outputs the orientation and length of a single fascicle from the mid-region of the image, and manual analysis involves a similar process. The averaging of multiple fascicles inevitably introduces some variability, because the number and location of fascicles detected by our method in any given image is not constant. It is therefore expected that the comparisons shown here would not yield identical results. Unfortunately, these comparisons cannot tell us how much of the differences are attributable to measurement errors, and how much is due to the detection of multiple fascicles. However, if desired, our method could be modified to only include one fascicle (e.g. the most central of those detected).

\subsection{Analysis of videos}The effect of detecting a variable number of fascicles per image is particularly evident when analysing videos. In general, a larger number of individual fascicles were detected in slower movements, and fewer fascicles in faster movements such as walking. This variable number of fascicles contributes to the variability of the mean traces shown in Figure ~\ref{fig:Fig6}. We believe that the mean value (or median) is probably not the best metric to use with this approach, and it may instead be preferable to include all detected fascicles in any subsequent analysis (and report a range of lengths), or to group fascicles from different regions of an image. Alternatively, the user could restrict fascicle detection to a narrower region, which would likely result in fewer fascicles being detected, but potentially result in a more representative mean value. 

When comparing the Ultratrack results with the mean fascicle length computed by our approach, the ICC values for the trials in Figure 6 varied widely, from 0.45 to 0.98. However, the use of the mean of all fascicles detected by our approach for comparison also had a large effect on the ICC values. In some frames, our method shows a bias toward detecting only longer or only shorter fascicles, resulting in a large jump in the mean signal, and in turn decreasing the ICC value when comparing to Ultratrack. We chose not to filter the mean traces produced by our approach in the interest of full disclosure, but also because it is not clear how to optimally filter a signal of this type whilst maintaining the important content. Other means of identifying an ‘average’ fascicle trajectory could include alternative methods of computing a mean (e.g. Lehmer mean), or some kind of filtering approach.

\begin{figure}[!h]
  \includegraphics[keepaspectratio=true, scale=0.6]{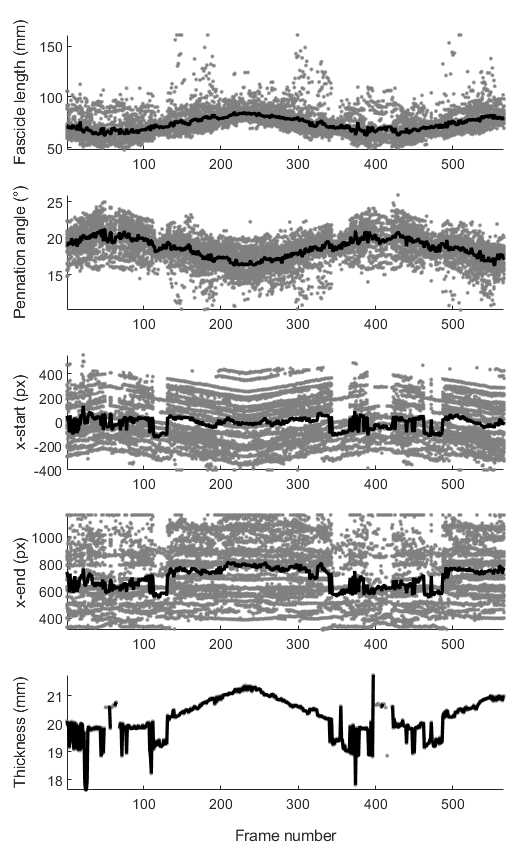}
  \caption{Example of all outputs that are stored in an xls file after analysing a video. A: fascicle length, B: pennation angle, C: x-axis location of each fascicle start point, D: x-axis location of each fascicle end point, E: thickness at the centre of the image. Mean traces are superimposed on each panel to make visualisation easier.}
  \label{fig:FigS4}
\end{figure}

In spite of the frame-to-frame variability in fascicle detection mentioned above, it is noteworthy that the results produced by Ultratrack always lie within the range of detected fascicle length values from our approach. Moreover, Ultratrack was designed to compute fascicle length from a single fascicle, the trajectory of which must be defined by the user in the first frame of the video. Our approach removes the need to subjectively identify the fascicle (and detects multiple fascicles), whilst also providing information about pennation angle and muscle thickness (see Figure ~\ref{fig:FigS4}). It should be noted that with Ultratrack, the fascicle start and end coordinates (and thus fascicle length) computed in any given frame are partly dependent on their coordinates in the previous frame. In other words, fascicle length is somewhat time-dependent with this method. On the contrary, our method treats each frame as being independent, which partly explains why fascicles are often detected from different regions in consecutive (and visually similar) frames. In the future, one possible extension of our method would be to incorporate information about fascicle lengths from previous frames using some kind of recurrent neural network, which would likely result in more stable mean length patterns. 

\subsection{Limitations and future work}Clearly, our approach detects fascicles from different regions in different frames, often in spite of the fact that two frames appear to be visually identical. This is naturally because of subtle differences in pixel values not detectable by our eyes, but also points to a limitation of this kind of supervised deep learning approach. This kind of algorithm essentially performs pattern recognition at scale. During the training process, the algorithm can only ‘learn’ patterns that are evident in the training dataset. Thus, when analysing new (previously unseen) images, the algorithm can easily be fooled into misclassifying, even by very subtle changes in pixel values. One solution to this problem is to add more labelled data during training, however it is not feasible to include labelled data that represent every possible variant of this kind of image. In the future, it is likely that different or complementary approaches, such as unsupervised learning \cite{Kamnitsas2017} or causative reasoning \cite{Sun1994}, will be needed to allow truly robust algorithms to be developed. It should be noted that in some frames, no fascicles were detected by our method. In cases where no fascicles are detected, simple gap-filling procedures could be used, especially when this only happens for 1 or 2 frames at a time. Nonetheless, when collecting data, potential users of our approach should be aware that if it is difficult or impossible for a human to identify the relevant structures from an image, it is unreasonable to expect a computer to be able to do so.

We offer our method not as a complete solution, but as a building block for future supervised approaches, or to help stimulate efforts that use other, more ‘intelligent’ approaches in the future. Supervised deep learning approaches are not actually ‘intelligent’ in the sense that they perform pattern recognition, but do not attach any meaning or context to those patterns. This means that their ability to make inferences on new data is often poor, especially when faced with new images that are quite different to those in the training dataset. In order to produce an algorithm that is truly intelligent, and can analyse images as robustly as a human, we may need a different, as yet non-existent approach (see for example \cite{Zador2019}).

\section{Conclusions}We present a fully automated, open source approach that provides fascicle length, pennation angle and muscle thickness, and works with single images and videos, as well as multiple superficial muscles. This method can be used by anyone with GPU access, or alternatively using the provided Google colaboratory version which includes free GPU access. It would also be straight-forward to expand this approach, e.g. by adding one’s own labelled training data, and using the supplied source code to train a new model. This method was designed for offline processing, but in theory, with sufficient hardware it could be possible to implement the approach in near real-time in order to provide online measures of muscle architecture.

%Here is how you add footnotes. \footnote{Sample of the first footnote.}

\bibliographystyle{unsrt}


\begin{thebibliography}{10}

\bibitem{Cronin2013}
NJ~Cronin and Glen Lichtwark.
\newblock {The use of ultrasound to study muscle–tendon function in human
  posture and locomotion}.
\newblock {\em Gait {\&} Posture}, 37:305--12, 2013.

\bibitem{Seynnes2015}
O.R. Seynnes, J.~Bojsen-M{\o}ller, K.~Albracht, A.~Arndt, N.J. Cronin,
  T.~Finni, and S.P. Magnusson.
\newblock {Ultrasound-based testing of tendon mechanical properties: A critical
  evaluation}.
\newblock {\em Journal of Applied Physiology}, 118(2), 2015.

\bibitem{Cronin2011c}
N~J Cronin, C~P Carty, R~S Barrett, and G~Lichtwark.
\newblock {Automatic tracking of medial gastrocnemius fascicle length during
  human locomotion.}
\newblock {\em Journal of applied physiology (Bethesda, Md. : 1985)},
  111(5):1491--6, nov 2011.

\bibitem{Rana2009}
Manku Rana, Ghassan Hamarneh, and James~M Wakeling.
\newblock {Automated tracking of muscle fascicle orientation in B-mode
  ultrasound images.}
\newblock {\em Journal of biomechanics}, 42(13):2068--73, sep 2009.

\bibitem{Marzilger2018}
Robert Marzilger, Kirsten Legerlotz, Chrystalla Panteli, Sebastian Bohm, and
  Adamantios Arampatzis.
\newblock {Reliability of a semi-automated algorithm for the vastus lateralis
  muscle architecture measurement based on ultrasound images}.
\newblock {\em European Journal of Applied Physiology}, 118(2):291--301, feb
  2018.

\bibitem{Drazan2019}
John~F. Drazan, Todd~J. Hullfish, and Josh~R. Baxter.
\newblock {An automatic fascicle tracking algorithm quantifying gastrocnemius
  architecture during maximal effort contractions}.
\newblock {\em PeerJ}, 2019(7), 2019.

\bibitem{Farris2016}
Dominic~James Farris and Glen~A. Lichtwark.
\newblock {UltraTrack: Software for semi-automated tracking of muscle fascicles
  in sequences of B-mode ultrasound images}.
\newblock {\em Computer Methods and Programs in Biomedicine}, 128:111--118, may
  2016.

\bibitem{Caresio2017}
Cristina Caresio, Massimo Salvi, Filippo Molinari, Kristen~M. Meiburger, and
  Marco~Alessandro Minetto.
\newblock {Fully Automated Muscle Ultrasound Analysis (MUSA): Robust and
  Accurate Muscle Thickness Measurement}.
\newblock {\em Ultrasound in Medicine and Biology}, 43(1):195--205, jan 2017.

\bibitem{Zhou2018}
Guang~Quan Zhou, Yi~Zhang, Ruo~Li Wang, Ping Zhou, Yong~Ping Zheng, Olga
  Tarassova, Anton Arndt, and Qiang Chen.
\newblock {Automatic myotendinous junction tracking in ultrasound images with
  phase-based segmentation}.
\newblock {\em BioMed Research International}, 2018, 2018.

\bibitem{Seynnes2020}
Olivier~R. Seynnes and Neil~J. Cronin.
\newblock {Simple Muscle Architecture Analysis (SMA): An ImageJ macro tool to
  automate measurements in B-mode ultrasound scans}.
\newblock {\em PLOS ONE}, 15(2):e0229034, feb 2020.

\bibitem{Darby2013}
John Darby, Baihua Li, Nicholas Costen, Ian Loram, and Emma Hodson-Tole.
\newblock {Estimating skeletal muscle fascicle curvature from B-mode ultrasound
  image sequences}.
\newblock {\em IEEE Transactions on Biomedical Engineering}, 60(7):1935--1945,
  2013.

\bibitem{Tajbakhsh2016}
Nima Tajbakhsh, Jae~Y. Shin, Suryakanth~R. Gurudu, R.~Todd Hurst,
  Christopher~B. Kendall, Michael~B. Gotway, and Jianming Liang.
\newblock {Convolutional Neural Networks for Medical Image Analysis: Full
  Training or Fine Tuning?}
\newblock {\em IEEE Transactions on Medical Imaging}, 35(5):1299--1312, may
  2016.

\bibitem{Milletari2016}
Fausto Milletari, Nassir Navab, and Seyed-Ahmad Ahmadi.
\newblock {V-Net: Fully Convolutional Neural Networks for Volumetric Medical
  Image Segmentation}.
\newblock {\em IEEE Xplore}, jun 2016.

\bibitem{Cunningham2018}
Ryan Cunningham, Mar{\'{i}}a S{\'{a}}nchez, Gregory May, and Ian Loram.
\newblock {Estimating Full Regional Skeletal Muscle Fibre Orientation from
  B-Mode Ultrasound Images Using Convolutional, Residual, and Deconvolutional
  Neural Networks}.
\newblock {\em Journal of Imaging}, 4(2):29, jan 2018.

\bibitem{Cunningham2020}
Ryan~J. Cunningham and Ian~D. Loram.
\newblock {Estimation of absolute states of human skeletal muscle via standard
  B-mode ultrasound imaging and deep convolutional neural networks}.
\newblock {\em Journal of the Royal Society Interface}, 17(162), jan 2020.

\bibitem{Ronneberger2015}
Olaf Ronneberger, Philipp Fischer, and Thomas Brox.
\newblock {U-Net: Convolutional Networks for Biomedical Image Segmentation}.
\newblock {\em arXiv}, pages 1--8, 2015.

\bibitem{Schindelin2012}
Johannes Schindelin, Ignacio Arganda-Carreras, Erwin Frise, Verena Kaynig, Mark
  Longair, Tobias Pietzsch, Stephan Preibisch, Curtis Rueden, Stephan Saalfeld,
  Benjamin Schmid, Jean~Yves Tinevez, Daniel~James White, Volker Hartenstein,
  Kevin Eliceiri, Pavel Tomancak, and Albert Cardona.
\newblock {Fiji: An open-source platform for biological-image analysis}, jul
  2012.

\bibitem{Gillett2013}
Jarred~G Gillett, Rod~S Barrett, and Glen~A Lichtwark.
\newblock {Reliability and accuracy of an automated tracking algorithm to
  measure controlled passive and active muscle fascicle length changes from
  ultrasound.}
\newblock {\em Computer methods in biomechanics and biomedical engineering},
  16(6):678--87, jan 2013.

\bibitem{Salarian2016}
A~Salarian.
\newblock {Intraclass Correlation Coefficient (ICC)}, 2016.

\bibitem{Kamnitsas2017}
Konstantinos Kamnitsas, Christian Baumgartner, Christian Ledig, Virginia
  Newcombe, Joanna Simpson, Andrew Kane, David Menon, Aditya Nori, Antonio
  Criminisi, Daniel Rueckert, and Ben Glocker.
\newblock {Unsupervised domain adaptation in brain lesion segmentation with
  adversarial networks}.
\newblock In {\em Lecture Notes in Computer Science (including subseries
  Lecture Notes in Artificial Intelligence and Lecture Notes in
  Bioinformatics)}, volume 10265 LNCS, pages 597--609. Springer Verlag, 2017.

\bibitem{Sun1994}
Ron Sun.
\newblock {A Neural Network Model of Causality}.
\newblock {\em IEEE Transactions on Neural Networks}, 5(4):604--611, 1994.

\bibitem{Zador2019}
Anthony~M. Zador.
\newblock {A critique of pure learning and what artificial neural networks can
  learn from animal brains}, dec 2019.

\end{thebibliography}

\end{document}